\documentclass[aps,prd,superscriptaddress,10pt,showpacs,notitlepage]{revtex4}
	\usepackage{graphicx}
	\usepackage{amsmath,amssymb,mathrsfs}
\usepackage{epsf}
\usepackage{epsfig}

\usepackage{color}

\usepackage{amsmath}

\usepackage{amssymb}

\numberwithin{equation}{section}

\usepackage{subfigure}

\newcommand{\be}{\begin{equation}}
\newcommand{\ee}{\end{equation}}
\newcommand{\bea}{\begin{eqnarray}}
\newcommand{\eea}{\end{eqnarray}}

\newcommand{\bb}{\bibitem}
 
\newcommand{\eqn}{\begin{eqnarray}}
\newcommand{\eqnx}{\end{eqnarray}}

\begin{document}
\title{On the spin excitation energy of the nucleon in the Skyrme model}

\author{C. Adam}
\affiliation{Departamento de F\'isica de Part\'iculas, Universidad de Santiago de Compostela and Instituto Galego de F\'isica de Altas Enerxias (IGFAE) E-15782 Santiago de Compostela, Spain}
\author{J. Sanchez-Guillen}
\affiliation{Departamento de F\'isica de Part\'iculas, Universidad de Santiago de Compostela and Instituto Galego de F\'isica de Altas Enerxias (IGFAE) E-15782 Santiago de Compostela, Spain}
\author{A. Wereszczynski}
\affiliation{Institute of Physics,  Jagiellonian University,
Lojasiewicza 11, Krak\'{o}w, Poland}

\begin{abstract}
In the Skyrme model of nucleons and nuclei, the spin excitation energy of the nucleon is traditionally calculated by a fit of the rigid rotor quantization of spin/isospin of the fundamental Skyrmion (the hedgehog) to the masses of the nucleon and the Delta resonance. The resulting, quite large spin excitation energy of the nucleon of about $ 73\, \mbox{MeV}$ is, however, rather difficult to reconcile with the small binding energies of physical nuclei, among other problems. Here we argue that a more reliable value for the spin excitation energy of the nucleon, compatible with many physical constraints, is about $ 16\, \mbox{MeV}$. The fit of the rigid rotor to the Delta, on the other hand, is problematic in any case, because it implies the use of a nonrelativistic method for a highly relativistic system.
\end{abstract}
\maketitle 

\section{Introduction}
The Skyrme model \cite{skyrme} and its generalizations \cite{omega-B}-\cite{BPS} are considered possible candidates for a low-energy effective field theory (EFT) for strong interaction physics \cite{thooft} which should, thus, be able to describe physical properties of hadrons and atomic nuclei \cite{adkins}-\cite{Sp2}. In this process, one important step consists in fitting the parameters (coupling constants) of the Skyrme model to certain physical quantities. In relation to these fits,
it has been noticed that there is a problem with  the spin energy from
the collective quantization of rigid nonrelativistic rotations. The
values for the nucleon higher spin resonances corrspond to unphysically
high velocities (see, e.g., \cite{HaKi} where a few very specific
relativistic corrections in the standard Skyrme model were proposed which, however, do
not improve the fit values, in general). The growing success of the Skyrme model for
nuclear physics, from light nuclei to nuclear matter and with generalized
Skyrme models, justifies a revision of this problem with the new
results, which provide a clearer picture. 
It is, thus, the main purpose of the present contribution to scrutinize the fits determining the spin excitation energy of the nucleon (proton and neutron) within the Skyrme model.
Most of our considerations hold for a rather general class of Skyrme models where, in addition to some obvious restrictions (Poincare invariance, and the existence of a standard hamiltonian), we only assume that, while chiral symmetry is broken (either spontaneously - zero pion mass, or explicitly, via a pion mass term), isospin remains a symmetry (i.e., the neutral an charged pions have the same mass). 
The lagrangian of the corresponding general Skyrme model reads
\begin{equation}
\label{lag}
\mathcal{L} \equiv \mathcal{L}_2+\mathcal{L}_4+\mathcal{L}_6+\mathcal{L}_0.
\end{equation}
Here, $\mathcal{L}_0$ is a potential term, and we assume
\be
\mathcal{L}_0 = -\lambda_0 \mathcal{U}(\mbox{Tr\;} (1-U)).
\ee
Further
 (the  $\lambda_i$ are non-negative (and, in general, dimensionfull) coupling constants)
\be
\mathcal{L}_2 = -\lambda_2 \mbox{Tr\;} L_\mu L^\mu, \;\; \mathcal{L}_4=\lambda_4 \mbox{Tr\;} [L_\mu, L_\nu]^2, \;\;
L_\mu = U^\dagger \partial_\mu U
\ee
are the sigma model and Skyrme terms, respectively.
Here $L_\mu = U^\dagger \partial_\mu U$ is the $su(2)$-valued left current, associated with
the $SU(2)$-valued  field $U = \sigma \cdot {\mathbb I} + i \pi^a \cdot \tau^a$, where
the four-vector of fields $(\sigma, \pi^a)$ is restricted to the surface
of the unit sphere, $\sigma^2+ \pi^a \cdot \pi^a =1$.
The sextic term is 
\be
\label{sextic}
\mathcal{L}_6=- \lambda_6 \mathcal{B}_\mu \mathcal{B}^\mu
\ee
where
\begin{equation}
\label{top-curr}
\mathcal{B}^\mu  = \frac{1}{24 \pi^2}\varepsilon^{\mu\nu\rho\sigma} \mbox{Tr\;} (L_\nu L_\rho L_\sigma) \, , \;\;\;
B=\int d^3 x \mathcal{B}^0
\end{equation}
is the topological current and $B$ is the baryon charge.

Proton and neutron are, thus, described by the same classical soliton (the hedgehog) and have the same mass (although a small mass difference may, in principle, be introduced after the semiclassical collective coordinate quantization has been performed \cite{bind}, \cite{Marl}).

Within the Skyrme model description of nucleons and nuclei, the mass of the nucleon (proton or neutron) $M_{\rm N}$ then receives two main contributions: the classical skyrmion (soliton) mass $M_0$ and the spin/isospin contribution $M_{\rm s}$ provided by the collective coordinate quantization of spin/isospin \cite{adkins}. Due to the rotational symmetry of the classical skyrmion (the hedgehog), spin and isospin are not independent, and only one of the two should be introduced as a set of collective coordinates. We shall choose the spin, for concreteness. If the rotating skyrmion is quantized as a rigid rotor, the spin contribution reads
\be
M_{\rm s} = \frac{\hbar^2 s(s+1)}{2I}
\ee
where $s$ is the spin quantum number ($s=\frac{1}{2}$ for the nucleon), and $I$ is the moment of inertia (the unique eigenvalue of the moment of inertia tensor). Further contributions to the mass can, in principle, be included into the Skyrme model, but are expected to be small for the nucleon. The two contributions should, therefore, sum up to the nucleon mass,
\be
M_{\rm N} = M_0 + M_{\rm s} = 938,9 \; {\rm MeV} 
\ee
(where our value is the average of proton and neutron mass).
An important question is, of course, which amount is contributed by each of the two terms.   Traditionally, the physical values for both the soliton mass and the spin energy are determined by fitting to the nucleon mass and to the Delta resonance, leading to a spin energy of about $M_{\rm s} = 73.2 \mbox{\; MeV}$ \cite{adkins}. We shall argue below that this fit is not reliable, because a non-relativistic method (the rigid rotor quantization) is used to describe a highly relativistic system (the Delta resonance). We shall, in fact, argue that a more likely realistic value is $M_{\rm s}\approx 16 \mbox{\; MeV}$.

\section{Values for $M_{\rm s}$ from nuclear physics}

To study this issue, let us consider the problem of higher nuclei and their binding energies. Within the Skyrme model, nuclei should be related to soliton solutions with higher baryon number $B$, which is then identified with the atomic weight number $A$, i.e., $A\equiv B$.
For a given baryon number, there may exist several soliton solutions $U_0^{(A,n)}$ with sufficiently close soliton masses $M_0^{(A,n)}$ (here $n$ labels the different solutions for a given $A$) such that they may give rise to stable or semi-stable nuclei. The rigid rotor quantization of spin and isospin will produce additional contributions $M_{{\rm s}}^{(A,n)}$ and $M_{{\rm i}}^{(A,n)}$ to the total mass, whose precise form depends on the symmetries and shapes of the corresponding soliton solutions. 
The leading contributions, however, may always be expressed like
\be \label{spin-ex-A}
M_{{\rm s}}^{(A,n)} = \frac{\hbar^2 s(s+1)}{2I_{{\rm s}}^{(A,n)}} + \ldots \; , \qquad M_{{\rm i}}^{(A,n)} = \frac{\hbar^2 i(i+1)}{2I_{{\rm i}}^{(A,n)}} + \ldots 
\ee
where the dots stand for subleading terms. Here, the moments of inertia $I_{{\rm s}}^{(A,n)}$ and $I_{{\rm i}}^{(A,n)}$ refer to the smallest eigenvalues of the corresponding tensors. The asymmetry (difference between different eigenvalues for a given soliton) will be small for approximately round (spherically symmetric) nuclei, but may become quite big for some skyrmions, especially for smaller $A$. As is true for all quantitative results, also the amount of asymmetry depends, of course, on the specific Skyrme model under consideration. In general, and particularly for sufficiently large $A$ and approximately round skyrmions, the moments of inertia grow with $A$ ($I_{{\rm s}}^{(A,n)}$ typically like $I_{{\rm s}}^{(A,n)} \sim A^\frac{5}{3}$, and $I_{{\rm i}}^{(A,n)}$ at least like $I_{{\rm i}}^{(A,n)} \sim A$). Further, $s$ is  not too big for physical nuclei even for large $A$, so the contribution of $M_{{\rm s}}^{(A,n)} $ to the masses of larger nuclei is usually very small. $i>i_3$, on the other hand, may take rather large values (for nuclei with a big imbalance between neutrons and protons - remember that 
$
 i_3 = \frac{1}{2} (Z-N)
$
is one-half the difference between protons and neutrons), so $M_{{\rm i}}^{(A,n)}$ may lead to appreciable contributions to the nuclear masses.
 
Further, for each solution $U_{0}^{(A,n)}$, there will exist a ground state (where the spin and isospin quantum numbers $s$ and $i$ take their minimum possible values) and excited states (higher values of $s$ and $i$). The identification of a skyrmion $U_{0}^{(A,n)}$ together with its quantum numbers $s$ and $i$ with a nucleus or nuclear state is, in general, a nontrivial problem, where both the symmetry of the solution and the quantum numbers of the nuclear state have to be taken into account (remember that $s$ and $i_3 \le i$ are good quantum numbers of nuclei). 

Once a nucleus X with mass $M_{\rm X}$ is identified with a soliton with mass 
\be
M^{(A,n)} = M_{0}^{(A,n)} + M_{{\rm s}}^{(A,n)} + M_{{\rm i}}^{(A,n)} ,
\ee
the corresponding binding energy within our approximation is
\be
E_{{\rm b}}^{(A,n)} = A M_{\rm N} - M^{(A,n)} = \left[ A M_0 - M_{0}^{(A,n)} \right] + \left[ A M_{\rm s} - M_{{\rm s}}^{(A,n)} - M_{{\rm i}}^{(A,n)} \right] 
\ee
where the classical (purely solitonic) binding energy is
\be
E_{{\rm b},0}^{(A,n)} =  A M_0 - M_{0}^{(A,n)}  .
\ee
Let us consider the helium nucleus as an example. Helium has $s=0$ and $i_3=0$, so it should be identified with a $B=4$ skyrmion with zero spin and isospin excitations, leading to the binding energy
\be
E_{{\rm b.He}} = 4 M_0 - M_{0,{\rm He}} + 4 M_{\rm s} .
\ee
Experimentally, the binding energy per baryon is $\frac{1}{4} E_{{\rm b.He}} = 7.2 \, {\rm MeV}$, so
\be
\frac{1}{4} \left( 4 M_0 -  M_{0,{\rm He}} \right)  +  M_{\rm s} = 7.2 \, {\rm MeV} .
\ee
Further conclusions depend on the value of the classical (skyrmion) binding energy $4 M_0 - M_{0,{\rm He}}$. In the original version of the Skyrme model, these binding energies are way too high (the binding energy per baryon number $M_0 - \frac{1}{4}M_{0,{\rm He}} \sim 100 \, {\rm MeV}$), so this relation cannot be satisfied. Recently, however, versions of the Skyrme model have been proposed which lead to very small or even exactly zero classical binding energies \cite{BPS}, \cite{BPSM}, \cite{bjarke1}, \cite{sutBPS}. If we assume zero classical binding energy for helium for the moment, this gives
$
M_{\rm s} = 7.2 \, {\rm MeV},
$
whereas for nonzero (positive) classical binding energies this seems to be an upper limit. 

We shall present arguments below which indicate that this value is too small and that a more realistic value should be 
\be
M_{\rm s} \approx 16 \, {\rm MeV}.
\ee
In a first instant, from our considerations above, this seems to imply that the classical binding energy of the skyrmion describing the helium nucleus should be negative, $M_{0,{\rm He}} >4 M_0$, i.e., the helium soliton should be classically unstable. We think that, instead, this just indicates that the description exposed above is incomplete, and a more complete description of, e.g., the helium nucleus within the Skyrme model 
requires the inclusion of further contributions to the mass which may, e.g., come from the quantization of further d.o.f. (e.g., vibrational modes). In some sense, these additional contributions should be related to "finite size effects" taking into account the smallness of the helium (and other small nuclei), as we shall argue below.   

The value $M_{\rm s} \approx 16 \, {\rm MeV}$ may be justified by the following arguments. The first argument only needs properties of the nucleon itself, and we shall call it the "self-consistent moment of inertia". It is based on the simple observation that for the nucleon, as described by the spherically symmetric hedgehog in the Skyrme model, there exists a relation between the moment of inertia $I$, the root-mean square nucleon radius $R_{\rm rms}$ and the classical skyrmion mass $M_0$. Indeed, the moment of inertia for a spherically symmetric mass (static energy) distribution is the same about any axis, and choosing, e.g., the $z$ axis we easily find
\be
I = \int d^3 x \rho_E (r) (x^2 + y^2) = \frac{8\pi}{3} \int dr r^4 \rho _E (r)
\ee
where $\rho_E (r)$ is the spherically symmetric energy density. The root-mean square (rms) radius, on the other hand, is defined as
\be
R_{\rm rms}^2 = \langle r^2 \rangle = \frac{\int d^3 x r^2 \rho_E (r)}{\int d^3 x \rho_E (r)} = \frac{4\pi \int dr r^4 \rho_E (r)}{M_0} =
\frac{3}{2}\frac{I}{M_0}
\ee
where $M_0 \equiv \int d^3 x \rho_E (r)$. This leads to the relation
\be
M_0 R_{\rm rms}^2 = \frac{3}{2}I
\ee
between $I$, $M_0$ and $R_{\rm rms}$, as announced. Replacing now $M_0$ and $M_{\rm s}$ by $I$ in the nucleon mass formula $M_{\rm N} = M_0 +M_{\rm s}$ leads to the self-consistent equation for the moment of inertia $I$
\be
M_{\rm N} = \frac{3}{2}\frac{I}{R_{\rm rms}^2} + \frac{3}{8}\frac{\hbar^2}{I}.
\ee
We prefer to replace $I$ by $M_{\rm s}$, leading to 
\be
M_{\rm N} = \frac{9\hbar^2}{16 R_{\rm rms}^2 } \frac{1}{M_{\rm s}} + M_{\rm s}
\ee
with the solution
\be
M_{\rm s} = \frac{1}{2}M_{\rm N} - \frac{1}{2}M_{\rm N} \sqrt{ 1- \left( \frac{3\hbar}{2M_{\rm N} R_{\rm rms}} \right)^2 } .
\ee
In particular, for the physical values $M_{\rm N} = 938.9 \, {\rm MeV}$, $R_{\rm rms} = 1.25 \, {\rm fm}$ (and $\hbar = 197.3 \, {\rm MeV \, fm}$), we get
\be
M_{\rm s} = 15.2 \, {\rm MeV} .
\ee

The remaining arguments relate the spin excitation energy $M_{\rm s}$ of the nucleon to more extended nuclear systems, concretely to infinite nuclear matter and to the binding energies of larger nuclei. In the case of infinite nuclear matter, we observe that the value $M_{\rm s} \approx 16 \, {\rm MeV}$  coincides with the binding energy per nucleon of infinite nuclear matter, 
\be
\bar E_{\rm b}^\infty \equiv \frac{E_{\rm b} ^\infty }{A} \approx 16 \, {\rm MeV} .
\ee 
Infinite nuclear matter is an idealised system of nuclear matter, where Coulomb energy contributions, surface effects and the difference between protons and neutrons are not considered (such that effectively only strong interaction effects are present). In our considerations above, we did not include Coulomb energy contributions, although their inclusion into the Skyrme model is, in principle, straight forward (the coupling of skyrmions to the electromagnetic field is exactly known). In practise, these Coulomb energy calculations are, nevertheless, quite cumbersome. The difference between neutrons and protons is taken into account, within our Skyrme model considerations, by the term $M_{\rm i}^{(A,n)}$ (here in the limit of large $A$), so we may ignore it by just skipping this term. Finally, surface effects may probably be related in our setting to classical binding energies in some sense, but should be absent in the limit of very large $A$. If we assume that, at least in the limit of very large $A$, the classical binding energies per baryon number are negligible, then we find $M_{\rm s} =  \bar E_{\rm b}^\infty \approx 16 \, {\rm MeV}$. If, instead, there remains a nonzero (positive) classical binding energy per baryon number in the limit of large $A$, then $\bar E_{\rm b}^\infty$ is an upper limit, i.e., 
$M_{\rm s} <  \bar E_{\rm b}^\infty \approx 16 \, {\rm MeV}$.

Our third argument follows from the comparison to the binding energies of the semi-empirical mass formula 
(Weizs\"acker formula) 
\begin{equation}
 E_{{\rm b},X}^{\rm W}(A,Z) = a_{\rm V} A - a_{\rm S} A^{2/3} - a_{\rm C} Z (Z-1) A^{-1/3} 
 - a_{\rm A} \frac{(A-2Z)^2}{A} + \delta (A,Z), 
\end{equation}
where
\begin{eqnarray}
&& \nonumber \\ &&
\delta(n,Z) = \left\{ \begin{array}{cl}
a_{\rm P} A^{-3/4} & N \; \textrm{and} \; Z \; \textrm{even}, \\
0 & A \; \textrm{odd}, \\
- a_{\rm P} A^{-3/4} & N \; \textrm{and} \; Z \; \textrm{odd},
\end{array} \right.
\nonumber \\ && \nonumber \\ &&
a_{\rm V} = 15.5 \; {\rm MeV}, \quad a_{\rm S} = 16.8 \; {\rm MeV}, \quad a_{\rm C} = 0.72 \; {\rm MeV},  \nonumber
\\ &&
a_{\rm A}= 23 \; {\rm MeV}, \quad a_{\rm P} = 34 \; {\rm MeV} .  \nonumber 
\end{eqnarray}
Here, the pairing term $a_{\rm P}$ is related to single-nucleon quantum properties and it is probably difficult to directly reproduce this term in a Skyrme-model context. As said, we did not include the Coulomb energy into our considerations, but this can be done and leads to a contribution which is very similar to the Coulomb contribution $a_{\rm C}$ in the  Weizs\"acker formula. Further, if we assume $i=i_3$ (the minimum possible value for the isospin energy), then our isospin term is similar to the asymmetry term $a_{\rm A}$. They are almost identical if $I_{\rm i}^{(A,n)} \sim A$. Finally, there is no term in our Skyrme-model binding energies which can be related to the surface term $a_{\rm S}$. It is not possible to relate this term to the classical (soliton) binding energies, because classical binding energies tend to increase the total binding energy, whereas the surface term reduces it. The surface term is particularly relevant for small nuclei like, e.g., the helium. A more complete mass calculation within the Skyrme model context should, therefore, produce a similar term, which is why we said earlier that some "finite size effects" should remove the tension between $M_{\rm s} \approx 16$ MeV and the helium binding energy.

Finally, there is no spin contribution in the Weizs\"acker formula, but, as said, these contributions are very small for larger $A$. If we now compare the remaining terms, we get
\be
  \left[ A M_0 - M_{0}^{(A,n)} \right] + A M_{\rm s} \approx a_{\rm V} A - a_{\rm S} A^{2/3} .
\ee
If we ignore the surface term (either by assuming that a similar term is produced on the l.h.s. by a more complete treatment of the Skyrme model, or by assuming a sufficiently large $A$), then we again find that $M_{\rm s} \approx a_{\rm V} \approx 16 \, {\rm MeV}$ for negligible classical (soliton) binding energies, or the upper bound $M_{\rm s} < a_{\rm V}$ for nonzero classical binding energies.

\section{The fit to the Delta}

Originally, the value of $M_{\rm s}$ was, instead, determined by a fit to the Delta resonance \cite{adkins}, and the resulting value is incompatible with the value of $M_{\rm s} \approx 16 \, {\rm MeV}$. Indeed, if the Delta resonance is interpreted as the $i=s=\frac{3}{2}$ excitation of the same $A=1$ skyrmion within the rigid rotor quantization, then the mass of the Delta resonance is
\be
M_\Delta = M_0 + \frac{\frac{3}{2}(\frac{3}{2}+1)\hbar^2}{2I} = M_0 + 5 M_{\rm s}
\ee
and, therefore,
\be
M_\Delta - M_{\rm N} = 4M_{\rm s} \qquad \Rightarrow \qquad M_{\rm s} = \frac{1}{4} \left( M_\Delta - M_{\rm N} \right) = 73.2 \, {\rm MeV}
\ee
where the Delta mass $M_\Delta = 1232 \, {\rm MeV}$ was used. This discrepancy, obviously, requires an explanation. 

Here we shall argue that the trustworthy value is $M_{\rm s} \approx 16 \, {\rm MeV}$, and that the much higher result from the rigid rotor quantization of the Delta resonance cannot be trusted. Our argument (but see also \cite{HaKi}, \cite{PSchZ}) is that the rigid rotor approximation is valid only in a non-relativistic regime. By re-interpreting the rotational excitations in the fit to the Delta as classical rotations we shall find, however, that these rotations are highly relativistic and the application of the rigid rotor approximation is, therefore, not justified. Indeed, for the nucleon spin excitation energy we have
\be
M_{\rm s} = \frac{3}{8}\frac{\hbar^2}{I} \qquad \Rightarrow \qquad I = \frac{3}{8}\frac{\hbar^2}{M_{\rm s}} .
\ee
In order to interpret this as a classical rotation, we now write
\be \label{Ms-omega-rel}
M_{\rm s} = \frac{1}{2} \omega_{\rm N}^2 I \qquad \Rightarrow \qquad \omega_{\rm N}^2 = \frac{2M_{\rm s}}{I} = \frac{16}{3}\frac{M_{\rm s}^2}{\hbar^2}
\ee
or finally 
\be
\omega_{\rm N} = \frac{4}{\sqrt{3}}\frac{M_{\rm s}}{\hbar}
\ee
where $\omega_{\rm N}$ is the classical circular frequency corresponding to the rotational excitation of the nucleon. Using the value $\hbar = 197.3 \, {\rm MeV \, fm}$, we find, for $M_{\rm s} = 16 \, {\rm MeV}$,
\be
\omega_{\rm N} = 0.187 \, {\rm fm}^{-1}
\ee
which, together with typical nucleon radii $R_{\rm N}$ of the order of 1 fm, leads to equatorial rotation velocities of the order of $0.2 $ (in our units the speed of light is equal to one). In a first instance, relativistic effects can, therefore, be expected to be sufficiently small such that a non-relativistic treatment is essentially justified. Still, depending on the stiffness of skyrmionic matter, at this value of the circular frequency a notable deformation of the rotating skyrmion may occur, which will then introduce a certain error in the rigid rotor results. The situation is better for nuclei for higher $A$, where the spin excitation energy is 
\be
M_{\rm s}^{(A)} = \frac{\hbar^2 s(s+1)}{2I_{\rm s}^{(A)}}
\ee
which, together with $I_{\rm s}^{(A)} \sim A^\frac{5}{3} I$, leads to 
\be
\omega_A \equiv \sqrt{s(s+1)} \frac{\hbar}{I_{\rm s}^{(A)}} \sim \sqrt{\frac{4s(s+1)}{3}}A^{-\frac{5}{3}}\omega_{\rm N} .
\ee
Further, nuclear radii grow like $R_A \sim A^\frac{1}{3} R_{\rm N}$, which leads to equatorial rotation velocities of the order of
\be
R_A \omega_A \sim \sqrt{s(s+1)} A^{-\frac{4}{3}} R_{\rm N} \omega_{\rm N}
\ee
and, therefore, to nonrelativistic velocities, in general.

If, instead,  we fit to the Delta resonance, that is,
for $M_{\rm s} = 73.2\, {\rm MeV}$, we find
\be
\omega_{\rm N} = 0.857 \, {\rm fm}^{-1}
\ee
and, therefore, equatorial rotation velocities which are firmly in the relativistic regime. Things get much worse for the circular frequency of the Delta. The moment of inertia $I$ remains the same, but the relation to the circular frequency now is
\be
5 M_{\rm s} = \frac{1}{2} \omega_\Delta^2 I \qquad \Rightarrow \qquad \omega_\Delta = \sqrt{5} \frac{4}{\sqrt{3}}\frac{M_{\rm s}}{\hbar}
\ee
which, for $M_{\rm s} = 73.2\, {\rm MeV}$, leads to
\be
\omega_\Delta = 1.916 \, {\rm fm}^{-1} .
\ee
This implies that, for all realistic nucleon radii, we are in an extremely relativistic regime. We conclude that the rigid rotor approximation for the Delta resonance is not reliable. The Delta resonance requires a fully relativistic treatment within the Skyrme model. We emphasize that we did not have to use specific calculations of particular Skyrme models to reach this conclusion. Arguments of a rather general, model-independent character are sufficient.

\section{Summary and Discussion}
It was the main purpose of the present contribution to point out that using the results of the rigid rotor quantization of spin/isospin for the $B=1$ skyrmion to fit the Delta resonance mass is probably not reliable, because the assumption of a "rigid rotor" in the rigid rotor quantization makes this a non-relativistic procedure, whereas the Delta is a highly relativistic "particle" (resonance). Further, we presented some arguments which favor a value for the spin excitation energy of the nucleon within the Skyrme model of about $M_{\rm s}\sim 16 \mbox{ MeV}$.
Before finishing, we want to add some observations.

Firstly, if, e.g., the standard Skyrme model (the one without $\mathcal{L}_6$, i.e., for $\lambda_6 =0$) is fitted to the properties of some nuclei instead of to the Delta, then much smaller values for $M_{\rm s}$ (i.e., much closer to $16\mbox{ MeV}$) may result. In the standard Skyrme model, traditionally the following parametrization is used,
\be
\lambda_2 = \frac{F_\pi^2}{16} \, ,\quad \lambda_4 = \frac{1}{32e^4}
\ee
and for the potential term one chooses the pion mass potential
\be
\mathcal{L}_0 = -\frac{1}{8}m_\pi^2 F_\pi^2 \mbox{Tr\,}(1-U).
\ee
Here, $m_\pi = 138\mbox{ MeV}$ is the pion mass which is assumed to take its physical value. Further, $F_\pi$ and $e$ are considered to be parameters of the model whose values should be determined by fits to physical quantities. The fit to the Delta leads to $F_\pi = 108 \mbox{ MeV}$, $e = 4.84$ and, of course, provides $M_{\rm s} = 73.2 \mbox{ MeV}$ by construction \cite{adkins}.

In \cite{wood1}, it was proposed to fit to properties of the lithium-6 nucleus. The resulting fit values are
$
F_\pi = 75.20 \mbox{ MeV} \; , \quad e = 3.263 
$
leading to $M_{\rm s} = 21.7 \mbox{ MeV}$ (see \cite{wood2}), which is much closer to the value of 16 MeV.

Secondly, one may, of course, wonder how trustworthy our proposed value of 16 MeV is. After all,  skyrmions are rather complicated objects with infinitely many degrees of freedom which can be excited, whereas in our simple discussion we only took their classical masses and spin/isospin excitations into account. Our main physical arguments, however, were based either on large nuclei or on infinite nuclear
matter, and in these large systems excitations provide rather small contributions to the total masses. Rotational excitation energies become small because the moments of inertia grow, and vibrational excitation energies become small because larger bodies composed of a given material have smaller fundamental vibrational frequencies than smaller bodies. Still, the value of 16 MeV should not be considered a precise prediction but just an indication about the correct value of $M_{\rm s}$.

Thirdly, there have been some attempts to improve on the description of the Delta by performing the quantization of the (iso-)spinning degrees of freedom about a classically isospinning skyrmion instead of a static one \cite{isospinning}. The common finding was that, for realistic (i.e., physically motivated) values of the parameters, stable isospinning skyrmions for the Delta cannot be found. But the Delta is a resonance with its (complex) mass pole embedded in the continuum above the threshold for the decay into a pion and a nucleon, so, with hindsight, these findings are not too surprising. A viable description of the Delta should both be fully relativistic and take into account its character as a resonance. 

Finally, we want to briefly comment on the spin excitation energies (\ref{spin-ex-A}) for large $A$.
As said, for large $A$ (large nuclei) they are small, but by considering different values of $s$ for a given nucleus they still seem to give rise to (rather narrow) spin excitation bands. Further, rotational bands are known to occur for some large atomic nuclei, so one might want to relate these to the spin excitations in the Skyrme model. Unfortunately, this is not possible for large nuclei, for the following reason. In large nuclei, rotational bands only may occur for spherically non-symmetric nuclei. Within the quantum mechanical $A$-particle (nucleon) description of a nucleus of atomic weight number $A$, a spherically symmetric rotating configuration cannot be distinguished from a static one. As a consequence, only the "hills" or "mountains" on the surface of a spherically non-symmetric nucleus may participate in the rotation and provide the moments of inertia, leading to much smaller moments of inertia and much bigger rotational excitation energies than a rigid body would. In the Skyrme model language, this means that a large nucleus should be considered an object which contains a sufficient amount of (almost) perfect fluid, such that the "mountains"  on the surface may follow a rotational movement without affecting the spherical core. Classically, the Skyrme model (in particular, in regions of rather high density) gets closer to a perfect fluid when the sextic term $\mathcal{L}_6$ is included \cite{BPS-fluid}. On the other hand, a quantum description for such an (almost) perfect fluid model is probably rather difficult to obtain (for a recent proposal for a quantum theory of fluids we refer to \cite{GriSut}).  For small nuclei, however, the rigid rotor quantization (eventually with the inclusion of some soft vibrations) has lead to very successful descriptions of excitational spectra (see, e.g., \cite{wood2}; for very recent results see \cite{lau} for $^{12}$C, or \cite{HalcrowKing2016} for $^{16}$O), so it seems that small nuclei may be described as essentially rigid bodies which tend to rotate as a whole, either without changing their shape at all, or by just vibrating between a small number of possible shapes (local minima of the energy functional).

\section*{Acknowledgements}
The authors acknowledge financial support from the Ministry of Education, Culture, and Sports, Spain (Grant No. FPA 2014-58-293-C2-1-P), the Xunta de Galicia (Grant No. INCITE09.296.035PR and Conselleria de Educacion), the Spanish Consolider-Ingenio 2010 Programme CPAN (CSD2007-00042), and FEDER.


\begin{thebibliography}{9}
\bibitem{skyrme} T. H. R. Skyrme, Proc. Roy. Soc. Lon. {\bf 260},
127 (1961); Nucl. Phys. {\bf 31}, 556 (1962); J. Math. Phys. {\bf
12}, 1735 (1971).
\bibitem{omega-B}  
A. Jackson, A.D. Jackson, A.S. Goldhaber, G.E. Brown, L.C. Castillejo, {\em Phys. Lett. B} {\bf 154} (1985) 101; 
M. Lacombe, L. Loiseau, R. Vinh Mau, W.N. Cottingham, {\em Phys. Lett. B} {\bf 161} (1985) 31; 
{\em Phys. Lett. B} {\bf 169} (1986) 121; 
U.-G. Meissner, I. Zahed, {\em Z. Phys. A} {\bf 327} (1987) 5;
 G. Holzwarth, R. Machleidt, {\em Phys. Rev. C} {\bf 55} (1997) 1088.
\bibitem{sextic-phen} 
A. Kanazawa, G. Momma, M. Haruyama, {\em Phys. Lett. B} {\bf 172} (1986) 403.
\bibitem{kop}
V.B. Kopeliovich, A.M. Shunderuk, G.K. Matushko,
{\em Phys. Atom. Nucl.} {\bf 69} (2006) 120. 
\bibitem{Ding2007} 
G.-J. Ding, M.-L. Yan, {\em Phys. Rev. C} {\bf 75} (2007) 034004.
\bibitem{sextic-skyrme} 
I. Floratos and B. Piette, {\em Phys. Rev. D} {\bf 64} (2001) 045009;
{\em J. Math. Phys.} {\bf 42} (2001) 5580.
\bibitem{bjarke2}
S.B. Gudnason, M. Nitta, Phys. Rev. D {\bf 91} (2015) 045027;  Phys. Rev. D {\bf 90}
(2014) 085007; Phys. Rev. D {\bf 89} (2014) 025012. 

\bibitem{BPSM} M. Gillard, D. Harland, M. Speight, {\em Nucl. Phys. B} {\bf 895} (2015) 272.
\bibitem{bjarke1}
S.B. Gudnason,  Phys. Rev. D {\bf 93} (2016), 065048; S.B. Gudnason, M. Nitta, arXiv:1606.02981.
\bibitem{BPS}
C. Adam, J. Sanchez-Guillen, A. Wereszczynski,
{\em Phys. Lett. B} {\bf 691} (2010) 105; 
C. Adam, J. Sanchez-Guillen, A. Wereszczynski,
{\em Phys. Rev. D} {\bf 82} (2010) 085015.
\bibitem{thooft} G. t'Hooft, Nucl. Phys. B{\bf 72}. 461 (1974); E. Witten, Nucl. Phys. B{\bf 160}, 57 (1979); E. Witten, Nucl. Phys. B{\bf 223}, 433 (1983).

\bibitem{adkins} G.S. Adkins, C.R. Nappi, E. Witten, {\em Nucl. Phys. 
B} {\bf 228} (1983) 552; G.S. Adkins, C.R. Nappi, {\em Nucl. Phys. B} {\bf 233} (1984) 109.
\bibitem{braaten}
E. Braaten, L. Carson,
{\em Phys. Rev. Lett.} {\bf 56} (1986) 1897; 
{\em Phys. Rev. D} {\bf 38} (1988) 3525.
\bibitem{carson}
L. Carson,
{\em Phys. Rev. Lett.} {\bf 66} (1991) 1406; 
L. Carson,
{\em Nucl. Phys. A} {\bf 535} (1991) 479;   
T.S. Walhout,
{\em Nucl. Phys. A} {\bf 531} (1991) 596. 
\bibitem{manton}
C.J. Houghton, N.S. Manton, P.M. Sutcliffe,
{\em Nucl. Phys. B} {\bf 510} (1998) 507;
R.A. Battye, P.M. Sutcliffe, {\em Nucl. Phys. B} {\bf 705} (2005) 384; R.A. Battye, P.M. Sutcliffe, {\em Phys. Rev. C} {\bf 73} (2006)  055205; 
R.A. Battye, N.S. Manton, P.M. Sutcliffe,
{\em Proc. Roy. Soc. Lond. A} {\bf 463} (2007) 261;
D.T.J. Feist, P.H.C. Lau, N.S. Manton, {\em Phys. Rev. D} {\bf 87} (2013) 085034.
\bb{wood1}
N.S. Manton, S.W. Wood, {\em Phys. Rev. D} {\bf 74} (2006) 125017.
\bibitem{wood2}
O.V. Manko, N.S. Manton, S.W. Wood,
{\em Phys. Rev. C} {\bf 76} (2007) 055203; 
R.A. Battye, N.S. Manton, P.M. Sutcliffe, 
S.W. Wood, {\em Phys. Rev. C} {\bf 80} (2009) 034323.
\bibitem{lau}
P.H.C. Lau, N.S. Manton, {\em Phys. Rev. D} {\bf 89} (2014) 125012;
{\em Phys. Rev. Lett.} {\bf 113} (2014) 23.
\bibitem{el-trans}
M. Haberichter, P.H.C. Lau, N.S. Manton, Phys. Rev. C {\bf 93} (2016) 034304.  
\bibitem{halcrow}
C.J. Halcrow, Nucl. Phys. B {\bf 904} (2016) 106.
\bb{HalcrowKing2016}
C.J. Halcrow, C. King, to be published.
\bb{harmonic_ap}  C. Adam, C. Naya, J. Sanchez-Guillen, A. Wereszczynski, Phys. Lett. B {\bf 726} (2013) 892.
\bibitem{bind}
C. Adam, C. Naya, J. Sanchez-Guillen, A. Wereszczynski, {\em Phys. Rev. Lett.} {\bf 111} (2013) 232501; {\em Phys. Rev. C} {\bf 88} (2013) 054313.
\bibitem{Marl} E. Bonenfant, L. Marleau, {\em Phys. Rev. D} {\bf 82} (2010) 054023;
E. Bonenfant, L. Harbour, L. Marleau, {\em Phys. Rev. D} {\bf 85} (2012) 114045; M.-O. Beaudoin, L. Marleau, {\em Nucl. Phys. B} {\bf 883} (2014) 328.
\bibitem{Sp2} J.M. Speight, {\em J. Geom. Phys.} {\bf 92} (2015) 30.
\bibitem{HaKi}
H. Hata, T. Kikuchi, Phys. Rev. D {\bf 82} (2010) 025017;
Progr. Theor. Phys. {\bf 125} (2011) 59.
\bibitem{sutBPS}
P. Sutcliffe,
{\em JHEP} {\bf 1008} (2010) 019;
{\em JHEP} {\bf 1104} (2011) 045;
{\em Mod. Phys. Lett. B} {\bf 29} (2015) 1540051.
\bibitem{PSchZ}
B. Piette, B. Schroers, W. Zakrzewski, Nucl. Phys. B {\bf 439} (1995) 205.
\bibitem{isospinning}
E. Braaten, J.P. Ralston, 
Phys. Rev. D {\bf 31} (1985)  598;
R. Rajaraman, H.M. Sommermann, J. Wambach, H.W. Wyld, 
Phys. Rev. D {\bf 33} (1986) 287;
R.A. Battye, S. Krusch, P.M. Sutcliffe, Phys. Lett. B {\bf 626} (2005) 126.
\bibitem{BPS-fluid}
C. Adam, C. Naya, J. Sanchez-Guillen, M. Speight, A. Wereszczynski, {\em Phys. Rev. D} {\bf 90} (2014) 045003; 
C. Adam, T. Kl\"ahn, C. Naya, J. Sanchez-Guillen, R. Vazquez, A. Wereszczynski,
{\em Phys. Rev. D} {\bf 91} (2015) 125037;
C. Adam, C. Naya, J. Sanchez-Guillen, R. Vazquez, A. Wereszczynski,
arXiv:1511.05160.
\bibitem{GriSut}
B. Gripaios, D. Sutherland,
Phys. Rev. Lett. {\bf 114} (2015) 071601. 


\end{thebibliography}
\end{document}